\documentclass[preprint]{rmaa}

\newcommand{\kms}   {km~s$^{-1}$}
\newcommand{\mjy}   {mJy~beam$^{-1}$}
\newcommand{\kl}    {k$\lambda$}
\newcommand{\lo}    {$L_{\sun}$}
\newcommand{\mo}    {$M_{\sun}$}
\newcommand{\h}     {H$_2$}
\newcommand{\hhs}   {HH~68-69}
\newcommand{\hhn}   {HH~94-95} 
\newcommand{\ngc}   {NGC~2264D} 
\newcommand{\mwc}   {MWC~1080} 
\newcommand{\vcy}   {V645~Cyg} 
\newcommand{\et}    {et al.}
\newcommand{\ie}    {i.\,e.,}
\newcommand{\eg}    {e.\,g.,}
\newcommand{\id}    {\it id.}
\newcommand{\rmaa}  {RevMexAA}

\newcommand{\ro}    {Rodr\'{\i}guez}

\newcommand{\xx}    {$\times$}
\newcommand{\D}[2]  {\mbox{#1$\arcdeg \pm$#2$\arcdeg$}}
\newcommand{\DS}[3] {\mbox{#1$\farcs$#2(#3)}}
\newcommand{\mmm}   {$\!\!\!\!$}
\newcommand{\mmt}   {$\!\!\!\!\!\!\!\!\!\!\!\!\!\!\!\!$}
\newcommand{\none}  {$...$}

\title{
Radio Continuum Observations towards Optical and Molecular Outflows}

\author{
Jos\'e M. Girart,
\affil{Departament d'Astronomia i Meteorologia, Universitat de Barcelona}
Salvador Curiel,
\affil{Instituto de Astronom{\'\i}a, UNAM, M\'exico D.F.}
Luis F. Rodr\'{\i}guez,
\affil{Instituto de Astronom{\'\i}a, UNAM, Morelia}
\and 
Jorge Cant\'o
\affil{Instituto de Astronom{\'\i}a, UNAM, M\'exico D.F.}
}

\fulladdresses{
\item{
Jos\'e M. Girart: Departament d'Astronomia i Meteorologia, 
Universitat de Barcelona, Av. Diagonal 647, 08028 Barcelona, Catalunya,
Spain; jgirart@am.ub.es. 
}
\item {
Luis F. Rodr\'{\i}guez: Instituto de Astronom\'{\i}a, 
Universidad Nacional Aut\'onoma de M\'exico, A. P. 3-72, 
(Xangari), 58089 Morelia, Michoac\'an, M\'exico 
(l.rodriguez@astrosmo.unam.mx)}
\item {
 Salvador Curiel and Jorge Cant\'o: Instituto de Astronom{\'\i}a, 
Universidad Nacional Aut\'onoma de M\'exico
Apdo. Postal 70-264, 04510 M\'exico D.F., M\'exico, (scuriel@astroscu.unam.mx)}
}

\shortauthor{Girart et al.}
\shorttitle{
Radio Continuum Observations towards Outflows}


\resumen{Presentamos observaciones de continuo en varias frecuencias,
realizadas con el VLA, en ocho regiones de formaci\'on estelar asociadas con
flujos moleculares y \'opticos: L1489, \hhs, \hhn, NGC~2264D, L1681B, L778,
MWC~1080 y V645~Cyg. Detectamos tres chorros radio t\'ermicos, L1489, YLW~16A
en L1681B y NGC~2264D VLA~7, asociados con flujos moleculares y/o flujos HH.
Los chorros t\'ermicos de radio en L1489 y NGC~2264D VLA~7 aparecen colimados
en la direcci\'on del flujo a mayor escala.  Presentamos la primera detecci\'on
tentativa de un chorro no t\'ermico de radio, L778 VLA~5, asociado con una
protoestrella de baja masa de clase I y con un flujo molecular. En \hhs, \hhn\
y en el flujo molecular en \ngc\ no hemos podido identificar las fuentes de
excitaci\'on de estos flujos. La emisi\'on de radio asociada con \vcy\ es
bastante extendida, $\sim 0.1$~pc y variable.  Detectamos tres radio fuentes en
la regi\'on de \mwc\ que podr\'{\i}an estar asociadas a fuentes j\'ovenes.}

\abstract{ We present multi-frequency VLA continuum observations towards 8
star  forming regions with molecular and optical outflows: L1489, \hhs, \hhn, 
\ngc, L1681B, L778, MWC~1080 and V645~Cyg.  We detect three thermal radio jets,
L1489, YLW~16A in L1681B and NGC~2264D VLA~7, associated with molecular and/or
HH outflows.  The L1489 and NGC~2264D VLA~7 thermal radio jets appear elongated
in the direction of the larger scale outflow.  We report the first tentative
detection of a non-thermal radio jet, L778 VLA~5, associated with a low mass
Class I protostar and powering a molecular outflow.  For \hhs, \hhn\ and the
molecular outflow in \ngc\ we could not identify a candidate of the exciting
source of these outflows.  The radio emission associated with \vcy\ is quite
extended, $\sim 0.1$~pc, and time variable.  We detect three radio sources in
the \mwc\ that could be associated with YSOs. }

\begin{document}
\maketitle

\section{Introduction\label{intro}}

%
\begin{table*}[t]
\scriptsize
\caption{Observed regions}
\label{obs}
\[
\begin{tabular}{lcccrcrrc}
\hline
\multicolumn{1}{c}{} &
\multicolumn{2}{c}{Phase Center} &
\multicolumn{1}{c}{} &
\multicolumn{1}{c}{} &
\multicolumn{1}{c}{} &
\multicolumn{2}{c}{Synthesized Beam} &
\multicolumn{1}{c}{rms}
\\
\multicolumn{1}{c}{} &
\multicolumn{2}{c}{\hrulefill} &
\multicolumn{1}{c}{$\lambda$} &
\multicolumn{1}{c}{} &
\multicolumn{1}{c}{Phase} &
\multicolumn{2}{c}{\hrulefill} &
\multicolumn{1}{c}{Noise}
\\
\multicolumn{1}{c}{Region} &
\multicolumn{1}{c}{$\alpha$(J2000)} &
\multicolumn{1}{c}{$\delta$(J2000)} &
\multicolumn{1}{c}{cm} &
\multicolumn{1}{c}{Date} &
\multicolumn{1}{c}{Calibrator} &
\multicolumn{1}{c}{HPFW} &
\multicolumn{1}{c}{PA} &
\multicolumn{1}{c}{$\mu$Jy~beam$^{-1}$}
\\
\hline
L1489 & 04 04 42.9 &  +26 18 56 &
  3.6 & 28/05/90   &  0403+260  & $0\farcs42\times0\farcs29$ & $-84\arcdeg$ & 15 \\
\hhs\ & 05 41 38.2 &$-$06 27 44 &
  3.6 & 07/07/95   &  0550+032  & $0\farcs33\times0\farcs25$ & $-19\arcdeg$ & 24 \\
\hhn\ & 05 43 39.0 &$-$02 35 09 &
    2 & 26/01/90   & 0541$-$056 & $6\farcs36\times4\farcs87$ & $-11\arcdeg$ & 63 \\
      &&&
  3.6 & 28/05/90   & 0541$-$056 & $0\farcs39\times0\farcs36$ & $+72\arcdeg$ & 15 \\
      &&&
  3.6 & 05/01/97   & 0539$-$057 & $0\farcs32\times0\farcs27$ & $+25\arcdeg$ & 10 \\
      &&&
    6 & 31/07/89   &  0550+032  & $6\farcs88\times3\farcs95$ & $-30\arcdeg$ & 34 \\
\ngc\ & 06 41 04.5 &  +09 36 20 &
    2 & 26/01/90   &  0725+144  & $5\farcs74\times4\farcs94$ & $+10\arcdeg$ & 59 \\
      &&&
  3.6 & 16/07/92   &  0629+104  &$10\farcs15\times8\farcs12$ & $ -9\arcdeg$ & 42 \\
      &&&
  3.6 & 07/07/95   &  0550+032  & $0\farcs30\times0\farcs25$ & $-45\arcdeg$ & 24 \\
      &&&
    6 & 31/07/89   &  0550+032  & $4\farcs74\times4\farcs16$ & $-34\arcdeg$ & 41 \\
L1681B& 16 27 28.0 &$-$24 39 33 &
  3.6 & 28/05/90   & 1626$-$298 & $0\farcs51\times0\farcs21$ & $+23\arcdeg$ & 17 \\
L778  & 19 26 28.9 &  +23 56 53 &
    2 & 26/01/90   &  1923+210  & $5\farcs27\times4\farcs89$ & $ -3\arcdeg$ & 57 \\
      &&&
  3.6 & 28/05/90   &  1923+210  & $0\farcs36\times0\farcs27$ & $+49\arcdeg$ & 16 \\
      &&&
    6 & 27/07/89   &  1923+210  & $4\farcs43\times3\farcs91$ & $-25\arcdeg$ & 33 \\
\vcy\ & 21 39 58.1 &   +50 14 20 &
    2 & 26/01/90   &  2200+420  & $5\farcs37\times4\farcs95$ & $+12\arcdeg$ & 66 \\
      &&&
  3.6 & 31/01/94   &  2146+608  & $11\farcs1\times10\farcs6$ & $-66\arcdeg$ & 30 \\
      &&&
    6 & 27/07/89   &  2200+420  & $4\farcs72\times4\farcs30$ & $+15\arcdeg$ & 32 \\
\mwc\ & 23 17 27.4 &   +60 50 48 &
    2 & 26/01/90   &  0014+612  & $5\farcs88\times4\farcs90$ &  $-2\arcdeg$ & 80 \\
      &&&
    6 & 27/07/89   &  2352+495  & $5\farcs50\times5\farcs03$ & $+55\arcdeg$ & 35 \\
\hline
\end{tabular}
\] 
\end{table*}


It is well established that in the early stages of star formation there is a
large mass loss in young stellar objects (YSOs). The two  most spectacular
manifestations of this phenomenon are the Herbig-Haro (HH thereafter) outflows
and the bipolar molecular outflows. HH outflows are shocks excited by highly
collimated, fast winds coming from YSO (\eg\ Hartigan \et\ 2000), while bipolar
molecular outflows are likely ambient gas swept up by those highly collimated
winds (\eg\ Richer \et\ 2000).  The energy sources of most molecular and HH
outflows are surrounded by large amounts of gas and dust, which contribute
significantly to their spectral energy distribution and produce such a large
extinction that YSOs are generally invisible at optical wavelengths (\eg\
Andr\'e 1997). Sometimes, they are so deeply embedded that in spite of the
recent developments in instrumentation at near-infrared wavelengths, they are
not easily detected even at these wavelengths (\eg\ Lada \& Lada 1991).
Therefore, observations in the mid and far infrared (and longward) wavelengths
are needed to identify the YSOs driving the molecular and HH outflows.
Submillimeter and millimeter observations are probably the most useful
techniques to identify YSOs, since they usually exhibit strong dust emission at
these wavelengths, and the mm interferometers can achieve high angular
resolution (\eg\ Wilner \& Lay 2000; \ro\ \et\ 1998).

\begin{table*}[t]
\scriptsize
\caption[]{Sources Detected from the Matching-Beam
Observations\tablenotemark{\lowercase{a}}}
\label{tmatch}
\[
\begin{tabular}{lcllccl}
\hline
\noalign{\smallskip}
\multicolumn{4}{c}{} &
\multicolumn{1}{c}{6 cm Flux} &
\multicolumn{1}{c}{2 cm Flux} &
\multicolumn{1}{c}{}
\\
\multicolumn{1}{l}{Region} &
\multicolumn{1}{l}{VLA} &
\multicolumn{1}{c}{\mmm$\alpha$(J2000)} &
\multicolumn{1}{c}{$\delta$(J2000)} &
\multicolumn{1}{c}{(mJy)} &
\multicolumn{1}{c}{(mJy)} &
\multicolumn{1}{l}{Identification}
\\
\noalign{\smallskip}
\hline
\noalign{\smallskip}
\hhn: 
&(1)& 05 43 28.36 &$-$02 35 26.3 & $0.50\pm0.09$ &$opb$&$bg$ \\
&(2)& 05 43 39.18 &$-$02 35 10.1 & $0.30\pm0.07$ & $0.39\pm0.12$ & \\
&(4)& 05 43 57.60 &$-$02 34 54.4 & $1.53\pm0.07$ &$opb$& \\
&(5)& 05 44 15.50 &$-$02 45 35.1 &\tablenotemark{b}&$opb$&$bg$ \\
&(6)& 05 44 47.19 &$-$02 35 06.1 &\tablenotemark{b}&$opb$& PMN J0544-0234?\\
\ngc:       
&(1)& 06 40 48.14 &  +09 33 05.0 &  $3.6\pm0.3$  &$opb$&$bg$ \\
&(2)& 06 40 50.47 &  +09 32 19.6 &  $0.9\pm0.2$  &$opb$&$bg$ \\
&(3)& 06 40 51.96 &  +09 31 54.0 & $18.6\pm0.3$  &$opb$&$bg$ \\
&(4)& 06 40 52.64 &  +09 29 54.0 & $20.0\pm0.5$  &$opb$&$bg$ \\
&(5)& 06 40 56.91 &  +09 38 42.1 & $2.78\pm0.08$ &$opb$&$bg$ \\
&(7)& 06 41 04.52 &  +09 36 20.5 & $0.72\pm0.04$ & $1.12\pm0.08$ & \\
&(12)&06 41 45.14 &  +09 47 03.0 &\tablenotemark{b}&$opb$ & 4C09.25 \\
L778:
&(1)& 19 26 02.69 &  +23 55 07.1 & $10.8\pm0.1$  &$opb$&$bg$ \\
&(2)& 19 26 02.76 &  +23 54 57.1 & $11.0\pm0.1$  &$opb$&$bg$ \\
&(3)& 19 26 12.64 &  +23 54 01.8 &  $1.0\pm0.1$  &$opb$&$bg$ \\
&(4)& 19 26 22.84 &  +23 53 46.8 &  $0.3\pm0.1$  &$opb$&$bg$ \\
&(5)& 19 26 28.77 &  +23 56 52.7 & $1.01\pm0.05$ & $0.22\pm0.06$ &IRAS~19243+2350 \\
&(6)& 19 26 29.15 &  +23 56 18.6 & $0.71\pm0.05$ & $0.79\pm0.09$ &    \\
&(7)& 19 26 30.11 &  +23 55 14.0 & $8.83\pm0.06$ & $2.66\pm0.24$ &$bg$ \\
V645~Cyg:  
&(1)& 21 39 22.60 &  +50 16 58.7 &  $8.2\pm0.2$  &$opb$&$bg$ \\
&(2)& 21 39 22.86 &  +50 10 24.7 &  $4.4\pm0.3$  &$opb$&$bg$ \\
&(3)& 21 39 32.92 &  +50 09 08.5 &  $3.1\pm0.2$  &$opb$& IRAS~21377+4955 \\
&(4)& 21 39 33.36 &  +50 09 10.8 &  $2.1\pm0.3$  &$opb$& \id\ \\
&(5)& 21 39 43.62 &  +50 15 16.8 & $1.72\pm0.06$ &$opb$&$bg$ \\
&(6)& 21 39 58.24 &  +50 14 21.5 & $0.57\pm0.04$ & $1.04\pm0.22$ & V645~Cyg \\
&(7)& 21 40 00.32 &  +50 06 51.6 &  $3.9\pm0.5$  &$opb$&$bg$ \\
&(8)& 21 40 00.87 &  +50 13 39.6 & $8.43\pm0.06$ & $2.89\pm0.15$ &$bg$ \\
MWC~1080:
&(1)& 23 16 46.83 &  +60 53 19.9 &  $4.1\pm0.2$  &$opb$&$bg$ \\
&(2)& 23 17 20.36 &  +60 48 20.8 & $0.43\pm0.08$ &$opb$&$bg$ \\
&(3)& 23 17 24.14 &  +60 50 44.9 & $0.17\pm0.04$ &$\la0.32$ & \\
&(4)& 23 17 25.52 &  +60 50 42.9 & $0.21\pm0.07$ &$\la0.32$ & MWC~1080? \\
&(5)& 23 17 27.43 &  +60 50 48.9 & $0.20\pm0.05$ &$\la0.32$ & \\
&(6)& 23 17 34.48 &  +60 56 43.1 &     $\sim18$  &$opb$&$bg$ \\
\hline
\end{tabular}
\] 
 $^a$ 
 ``$opb$'' indicates sources outside of the 2~cm primary beam. 
 ``$bg$'' means a likely background source. \\
 $^b$ Sources outside the primary beam at 6~cm. \\
\end{table*}

An alternative way to identify this type of sources is through interferometric 
radio continuum observations at centimeter wavelengths, carried out mainly with
the VLA (and more recently with MERLIN and the Australia Telescope). The VLA
allows to map large regions (e.g. the primary beam at 6~cm is  9$'$) with a
very high sensitivity. In addition, the spectral indices of the  sources can be
measured from multi-frequency radio continuum observations  and, therefore, it
allows to elucidate the nature of the radio emission. Combining this type of
observations with other criteria, such as the source coinciding with the
geometrical center of the outflow and its association with an infrared and/or
millimeter counterpart, has proven  to be a very useful tool to discriminate
among the candidates of the energy  source of the outflow (\eg\ HH~1-2: Pravdo
\et\ 1985; L1448:  Curiel \et\ 1990; L1287: Anglada \et\ 1994).   Recently, a
number of surveys at centimeter wavelengths have been  carried out in order
to identify the powering sources of molecular outflows (\eg\ Anglada \et\ 1992,
1998;  Beltr\'an 2001) and of HH objects (\eg\ \ro\ \& Reipurth  1994, 1998;
Avila, \ro, \& Curiel 2001).

For wavelengths longer than $\sim$1~cm, the continuum emission of the energy 
sources of molecular and HH outflows is often dominated by free-free emission 
from partially ionized outflows (\eg\ Anglada 1996).  These cm radio
observations allow to determine with great accuracy the position of the
exciting source and to determine the morphology and other physical parameters
of the ionized gas at small angular scales.  Thus, the  radio continuum sources
associated with YSOs, powering molecular and/or HH outflows, are usually found
to have the following characteristics (\eg\  Anglada 1996, \ro\ 1997): {\it
(1)} Relatively weak flux  densities in the cm regime (around 1 mJy or less);
{\it (2)} Spectral indices  that are flat or rise slowly with frequency
(typically between -0.1 and 1);  {\it (3)} No evidence of large time
variability; {\it (4)} No evidence of  polarization; and {\it (5)} in some of
the best studied cases they exhibit a  jet-like morphology, with their
orientation, in most cases, along the  molecular or HH outflow direction. 
Because of these properties, these objects  are known in the literature as
``thermal radio jets''.

In this paper we present matching--beam VLA observations at 2 and 6~cm and
sub-arcsecond angular resolution observations towards several star forming 
regions with associated molecular outflows and/or  Herbig-Haro objects. Most of
the selected regions were previously observed with  the VLA at only one
wavelength and with lower angular resolution.

\section{Observations}

The radio continuum observations were carried out towards 8 fields with the 
Very Large Array (VLA) of the National Radio Astronomy 
Observatory\footnote{NRAO is a facility of the National Science Foundation 
operated under cooperative agreement by Associated Universities, Inc.}\ 
between 1989 and 1997.  Matching--beam observations at 2 and 6~cm toward \hhn, 
\ngc, L778, \vcy\ and \mwc\ were carried out with the D and B/C configurations
respectively, which provided an angular resolution of $\sim$ 5$''$.  The 2 and
6~cm observations were done with a difference of about half a year.  The 
similar angular  resolution and the proximity in time of the observations allow
a good  estimation of the spectral indices.  In addition, sensitive 
subarcsecond angular resolution observations at 3.6~cm toward L1489, \hhs, 
\hhn, \ngc, L1681B and L778 were carried out with the A/B and A
configurations.  \ngc\ and \vcy\ were also observed at 3.6~cm in the D array. 
The absolute  amplitude calibrator was always 3C286.  The data were edited and
calibrated following the standard VLA procedures with the AIPS software
package. In \ngc\, a strong source was detected at 6~cm, outside the primary
beam response (HPFW) (at $\sim10'$E and $11'$N from the phase center). This
source was removed from the $u,v$ data following the recommended procedure (map
done using UVMAP with a shift of $x=-596''$,  $y=648''$, cleaned with APCLN and
subtracting the cleaned components from the $u,v$ data using UVSUB).  
Self-calibration was performed on those fields with strong sources 
($\ga$5~mJy): L778, V645~Cyg and NGC~2264D at 6~cm.  Final maps were done using
robust weighting of $\sim 0.5$ for the subarcsecond 3.6~cm observations and
using natural weighting  for the rest of the observations.  All the fluxes were
corrected by the VLA antenna primary beam response.  In Table~\ref{obs} we list
the observed regions, the phase center, the phase  calibrators, the synthesized
beam and the rms noise achieved for all the fields. The measured positions and
fluxes of the sources detected with the 2 and 6~cm  matching-beam observations
are given in Table~\ref{tmatch}. The positions and flux densities of the 3.6 cm
subarcsecond and low angular resolution observations are given in
Table~\ref{thigh} and \ref{tlow}, respectively. The fluxes and spectral indices
of the candidates for driving source, obtained from the matching--beam
observations, are shown in Table~\ref{tsize}. The deconvolved size, the
position angle and the projected physical size of the putative driving sources
that were spatially resolved at 3.6 cm are also shown in Table~\ref{tsize}.

\begin{table}
\scriptsize
\caption[]{Sources Detected at 3.6~cm (A and A/B array)}
\label{thigh}
\begin{tabular}{lllcc}
\hline
\noalign{\smallskip}
\multicolumn{1}{c}{Object} &
\multicolumn{2}{c}{} &
\multicolumn{1}{c}{\mmm Flux} &
\multicolumn{1}{c}{}
\\
\multicolumn{1}{l}{} &
\multicolumn{1}{c}{\mmt$\alpha$(J2000)} &
\multicolumn{1}{c}{\mmm$\delta$(J2000)} &
\multicolumn{1}{c}{\mmm (mJy)} &
\multicolumn{1}{c}{\mmm Identification\tablenotemark{a}}
\\
\noalign{\smallskip}
\hline
\noalign{\smallskip}
\multicolumn{2}{l}{L1489:}        && \\
1 &\mmt04 04 43.088&\mmm  +26 18 56.76 &\mmm$0.49\pm0.02$ &\mmm IRAS~04016+2610 \\
\multicolumn{2}{l}{\hhs:}         && \\
1 &\mmt05 41 45.802&\mmm$-$06 27 05.78 &\mmm$1.30\pm0.05$ &\mmm \\
\multicolumn{2}{l}{\hhn:}         && \\
1 &\mmt05 43 28.337&\mmm$-$02 35 25.88 &\mmm$0.60\pm0.05$ &\mmm $bg$ \\
2 &\mmt05 43 39.168&\mmm$-$02 35 09.51 &\mmm$0.29\pm0.02$ &\mmm \\
3 &\mmt05 43 39.295&\mmm$-$02 35 36.39 &\mmm$0.12\pm0.02$ &\mmm \\
\multicolumn{2}{l}{\ngc:}         && \\
5 &\mmt06 40 56.905&\mmm  +09 38 42.1  &\mmm $0.9\pm0.1$  &\mmm $bg$ \\
7 &\mmt06 41 04.516&\mmm  +09 36 20.52 &\mmm$0.62\pm0.03$ &\mmm \\
8 &\mmt06 41 06.503&\mmm  +09 36 05.28 &\mmm$1.49\pm0.06$ &\mmm \\
9 &\mmt06 41 06.579&\mmm  +09 34 32.15 &\mmm$0.21\pm0.06$ &\mmm \\
10&\mmt06 41 06.665&\mmm  +09 35 30.07 &\mmm$0.20\pm0.05$ &\mmm \\
\multicolumn{2}{l}{L1681B:}       && \\
1 &\mmt16 27 26.896&\mmm$-$24 40 49.79 &\mmm$1.80\pm0.03$ &\mmm YLW~15 \\
2 &\mmt16 27 26.921&\mmm$-$24 40 50.23 &\mmm$0.79\pm0.03$ &\mmm \id \\
3 &\mmt16 27 27.997&\mmm$-$24 39 32.87 &\mmm$0.67\pm0.03$ &\mmm YLW~16A \\
\multicolumn{2}{l}{L778:}         && \\
5 &\mmt19 26 28.786&\mmm  +23 56 53.39 &\mmm$0.69\pm0.02$ &\mmm IRAS 19243+2350 \\
6 &\mmt19 26 29.143&\mmm  +23 56 18.76 &\mmm$1.06\pm0.02$ &\mmm \\
7 &\mmt19 26 30.109&\mmm  +23 55 14.36 &\mmm$5.29\pm0.03$ &\mmm $bg$ \\
\noalign{\smallskip}
\hline
\end{tabular}
 $^a$ $bg$: likely a background source. \\
\end{table}

\begin{table}
\scriptsize
\caption[]{Sources Detected at 3.6~cm (D array)}
\label{tlow}
\begin{tabular}{lllcc}
\hline
\noalign{\smallskip}
\multicolumn{1}{c}{Object} &
\multicolumn{2}{c}{} &
\multicolumn{1}{c}{\mmm Flux} &
\multicolumn{1}{c}{}
\\
\multicolumn{1}{l}{} &
\multicolumn{1}{c}{\mmt$\alpha$(J2000)} &
\multicolumn{1}{c}{\mmm$\delta$(J2000)} &
\multicolumn{1}{c}{\mmm (mJy)} &
\multicolumn{1}{c}{\mmm Identification\tablenotemark{a}}
\\
\noalign{\smallskip}
\hline
\noalign{\smallskip}
\multicolumn{2}{l}{\ngc:}      && \\
5 &\mmt 06 40 56.91 &\mmm +09 38 42.1 &\mmm  $2.0\pm0.2$ &\mmm $bg$ \\
6 &\mmt 06 41 02.82 &\mmm +09 36 16.9 &\mmm $0.32\pm0.04$&\mmm IRAS~06328+0939 \\
7 &\mmt 06 41 04.52 &\mmm +09 36 20.5 &\mmm $1.17\pm0.07$&\mmm  \\
8 &\mmt 06 41 06.50 &\mmm +09 36 05.3 &\mmm    $\sim0.13$&\mmm  \\
11&\mmt 06 41 06.67 &\mmm +09 35 54.9 &\mmm $0.26\pm0.04$&\mmm  \\
\multicolumn{2}{l}{V645~Cyg:}  && \\
5 &\mmt 21 39 43.54 &\mmm +50 15 16.7 &\mmm $0.81\pm0.14$&\mmm $bg$ \\
6 &\mmt 21 39 58.21 &\mmm +50 14 20.9 &\mmm $0.99\pm0.03$&\mmm V645~Cyg \\
8 &\mmt 21 40 00.82 &\mmm +50 13 39.6 &\mmm $5.57\pm0.05$&\mmm $bg$ \\
\noalign{\smallskip}
\hline
\end{tabular}
 $^a$ $bg$: likely a background source. \\
\end{table}

Compared with the 2 and 3.6~cm maps, the 6~cm maps are more affected by the
presence of background sources due to the larger primary beam ($\sim 9'$) and
the increasing flux density with wavelength of these non-thermal sources.  In
order to identify them we have defined possible background  sources as those
which lie farther than $2'$ from the phase center (and therefore far away from
the geometrical center of the molecular or  optical outflow), and those which
exhibit a clearly non-thermal emission  (\ie\ a negative spectral index). In
addition, we looked for 20~cm  counterparts using the NVSS (NRAO VLA Sky
Survey) catalog (Condon \et\  1998) since non-thermal sources are stronger at
this wavelength.  We note that with these criteria, we do not expect to
identify all the  non-thermal sources within the 6 cm primary beam, since we
cannot determine  a spectral index for all the sources due to the significantly
lower  sensitivity of the 20~cm maps. The sources that meet these criteria are 
labeled as background, except for those which have a tentative identification 
at other wavelengths  (see Tables~\ref{tmatch}, \ref{thigh} and \ref{tlow}).  
We detected a total of 30 radio continuum sources in the 5 fields observed at 
6~cm with a peak flux above the 5-$\sigma$ level, and 21 of them are
identified  as background sources.  The number of background sources expected
within the  VLA primary beam at 6~cm can be estimated from the formulation
given by Anglada \et\ (1998):  
\begin{equation} <N> \simeq 1.1 \left( {S_0}/{mJy}
\right) ^{-0.75} 
\end{equation} 
For our typical detection threshold (0.18~mJy at 5-$\sigma$) we expect to
detect about 20 sources for the five fields, which is in agreement with our
initial estimate of 21 background sources.

Since the thermal radio jets are typically weak sources, in order to elucidate 
whether they are resolved at subarcsecond angular resolutions, we use $S_{\rm
total} - S_{\rm peak} \ga 2 \times \left(\Delta S_{\rm total} +  \Delta S_{\rm
peak} \right)$ as a criteria, where $S_{\rm total}$ and $S_{\rm peak}$ are the
total flux density and the peak intensity respectively, and $\Delta S_{\rm
total}$ and $\Delta S_{\rm peak}$ are their respective rms noises.  For those
sources with adequate signal-to-noise ratio ($SNR\ga8$) least-square fits of
Gaussian ellipsoids were done to obtain the deconvolved size and the position 
angle (PA) of the source. We have taken into account the beam smearing that
affects the sources which are far from the phase center, producing an apparent
elongation in the direction toward the phase center of approximately
$[\frac{d}{''}] \times \frac{\Delta \nu}{\nu}$, where d is the distance to the 
phase center, $\nu$ and $\Delta \nu$ are the observed frequency and bandwidth,
respectively .

\section{Results and Discussion}

\subsection{L1489}

L1489 is a well studied dark cloud located in the Taurus molecular cloud 
complex (\eg\ Myers \et\ 1988; Zhou \et\ 1989;  Miyawaki \& Hayashi 1992). 
Embedded in the dark cloud there is  a Class I YSO with a luminosity of
3.7~\lo, surrounded by an optical and near-IR nebulosity (\eg\ Myers \et\ 1987;
Zinnecker \et\  1992; Heyer \et\ 1990).  High angular  resolution observations
show a 2000~AU rotating circumstellar envelope around the YSO (Ohashi \et\
1996; Hogerheijde \et\  1998; Hogerheijde \& Sandell 2000).  HST observations
(Padgett \et\ 1999;  Wood \et\ 2001) revealed a dark dust lane elongated over
the  east-west direction ($PA=85\arcdeg$) with a unipolar nebula with its
symmetry axis perpendicular to the dust lane ($PA=175\arcdeg$), possibly
engulfing an embedded binary system. Associated with this source, there is a
low-velocity molecular outflow without a well defined bipolar morphology (Myers
\et\ 1992), although higher angular resolution observations reveal a weak and
compact bipolar component of the molecular outflow coinciding with the YSO and
oriented nearly in the north-south direction ($PA=165\arcdeg$), \ie\
perpendicular to the circumstellar envelope  (Hogerheijde \et\ 1998). Radio
continuum emission arising from the energy source of the L1489 molecular
outflow shows that the emission has a positive spectral index, $0.3\pm0.2$, and
is partially resolved in the NW-SE direction at subarcsecond scales (\ro\ 
\et\ 1989; Lucas \et\ 2000).  Optical and near-IR emission show a HH system 
(HH~360 through 362), that also appears to be powered by the same source 
(G\'omez, Whitney \& Kenyon 1997; Lucas \et\ 2000).

\begin{table*}[t]
\scriptsize
\caption[]{Candidates of powering Molecular/HH outflows}
\label{tsize}
\[
\begin{tabular}{lcrrcrrl}
\hline
\multicolumn{2}{l}{} &
\multicolumn{5}{c}{VLA Candidate Properties} &
\multicolumn{1}{l}{} 
\\
\cline{3-7}
\multicolumn{3}{l}{} &
\multicolumn{1}{r}{\mmm Spectral} &
\multicolumn{1}{c}{Deconvolved} &
\multicolumn{1}{c}{} &
\multicolumn{1}{c}{Size} &
\multicolumn{1}{c}{} 
\\
\multicolumn{1}{l}{Region} &
\multicolumn{1}{c}{\mmm Outflow} &
\multicolumn{1}{c}{Source} &
\multicolumn{1}{c}{Index} &
\multicolumn{1}{c}{Size$^a$} &
\multicolumn{1}{c}{\mmm PA} &
\multicolumn{1}{c}{(AU)} &
\multicolumn{1}{c}{Comments}
\\
\hline
L1489    &\mmm$\!\!\!\!$ CO outflow, HH 360& VLA 1               
 & \mmm 0.3$\pm$0.1 &\DS{0}{27}{2}\xx\DS{0}{15}{2} &\mmm \D{168}{7}& 38 
 & Class I YSO    \\
\hhn     & \hhn              & VLA 2               
 & \mmm 0.0$\pm$0.3 & $\la 0\farcs1$\xx$\la 0\farcs1$ &\mmt \none & $\la50$
 & Dustless object    \\
         &                   & VLA 3               
 &\mmm $-$0.4 to 1.3& $\la 0\farcs3$\xx$\la 0\farcs3$ &\mmt \none & $\la150$
 & Unknown counterpart \\
\ngc     &\mmm HH 125/225/226    & VLA 7               
 & \mmm0.4$\pm$0.1  &\DS{0}{46}{2}\xx\DS{0}{27}{2} &\mmm \D{162}{5}& 368 
 & Unknown counterpart \\
YLW 16A  &\mmm CO outflow       & VLA 3               
 & \mmm0.1$\pm$0.1  &\DS{0}{27}{2}\xx\DS{0}{13}{2} &\mmm \D{47}{7} & 43
 & Class I YSO    \\
L778     &\mmm CO outflow        & VLA 5               
 & \mmm-0.82$\pm$0.04 &\DS{0}{34}{1}\xx\DS{0}{09}{2} &\mmm \D{97}{2} & 142
 & Class I YSO?    \\
V645 Cyg &\mmm CO outflow        & VLA 6               
 & \mmm0.5$\pm$0.2  &\DS{6}{5}{4}\xx\DS{3}{6}{4}   &\mmm \D{7}{6}  & 23000
 & Herbig Ae/Be star     \\
MWC 1080 &\mmm HH/CO outflow        & VLA 4 
 & \mmm $\la$0.4    & $\la 5''$\xx$\la 5''$  & \none      & $\la13000$
 & Associated with MWC 1080 ? \\ 
MWC 1080 &\mmm HH/CO outflow        & VLA 3 
 & \mmm $\la$0.6    & $\la 5''$\xx$\la 5''$  & \none      & $\la13000$
 & Dusty object \\ 
MWC 1080 &\mmm HH/CO outflow        & VLA 5 
 & \mmm $\la$0.4    & $\la 5''$\xx$\la 5''$  & \none      & $\la13000$
 & Dusty object \\ 
\hline
\\
\end{tabular}
\]
\begin{list}{}{}
\item{$^{a}$} The value within the parenthesis gives the error in the last digit
\end{list}
\end{table*}

\begin{figure}
\includegraphics[width=\columnwidth]{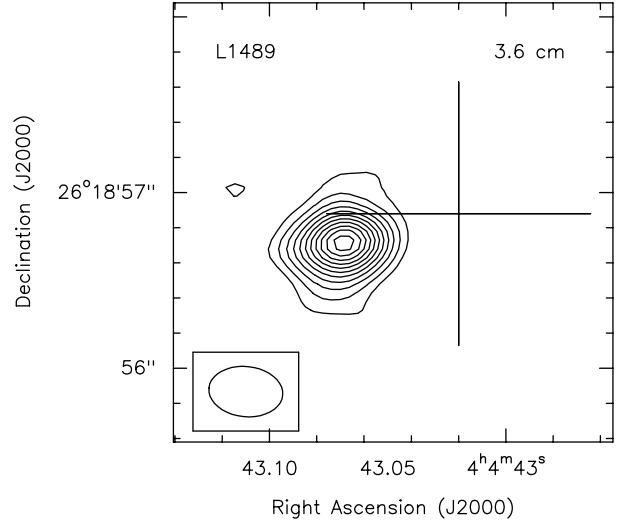}
\caption
{Subarcsecond angular resolution map at 3.6~cm of L1489.  Contours are $-$3,
3, 5, 7, 9, ... and 23 times the rms noise of the map, 15~$\mu$Jy~beam$^{-1}$.
The half power contour of the synthesized beam is shown in the bottom left
corner. The cross marks the position of the molecular disk around the YSO
(Hogerheijde \et\  1998).
\label{l1489}
}
\end{figure}

For this source we carried out subarcsecond angular resolution observations at
3.6~cm.  The map obtained is shown in Figure~\ref{l1489}.  The emission is
partially resolved (see Fig.~\ref{l1489}), mainly in the north-south 
direction, with a peak intensity of $0.35\pm0.01$~\mjy.  The deconvolved size
from a Gaussian fit is $0\farcs26\pm0\farcs01 \times 0\farcs13\pm0\farcs02$
and the position angle of the major axis is  $PA=171\arcdeg \pm5\arcdeg$. The
intensity and size measured are in agreement with the values obtained by Lucas
\et\ (2000).  The emission is extended along the outflow axis, defined by the
main axis of the compact component of the molecular outflow axis and the line
joining the HH~360 knots (Hogerheijde \et\  1998; G\'omez \et\ 1997), and
perpendicular to the molecular and dust circumstellar core that surrounds this
YSO (Ohashi \et\ 1996; Hogerheijde \et\ 1998).  We note, however, that HH~361
and HH~362, and the more extended molecular outflow do  not line up in the
north-south direction (see Fig.~1 from G\'omez \et\  1997).  This might be due
to a change of the outflow direction,  or to the presence of a second outflow
in the region.  Indeed, Lucas \et\ (2000) shows that the near-IR \h\ emission
traces a quadrupolar outflow. 

In order to compare the flux density with previous VLA observations, we made
natural  weighting maps by applying a Gaussian taper to the visibilities, which
provided  a $0\farcs5$ synthesized beam.  The total flux density measured from
this map  is $0.52\pm0.02$~mJy, which is in agreement with the lower angular 
resolution measurements at 2 and 6~cm by \ro\ \et\ (1989).   Combining flux
densities obtained at the three wavelengths, we obtain that the spectral index
is $0.3\pm0.1$, which is consistent with partially thick free-free emission.

\subsection{\hhs}

The \hhs\ objects were discovered by Reipurth \& Graham (1988), although this
is a poorly studied region.  The optical outflow is located in the remains of a
molecular cloud, which is probably in the process of destruction due to an
expanding HII region (Ogura \& Sugitani 1998). Detailed analysis of the IRAS
images led Cohen (1990) to suggest that \hhs\ are two independent HH systems,
powered by two infrared sources, IRAS~05391$-$0627C and IRAS~05393$-$0632. \ro\
and Reipurth (1994) and Avila \et\ (2001) failed to detect the radio continuum
emission towards the infrared sources.  Yet, Avila \et\ (2001) detected
emission towards HH~68b with a flux density of 0.18~mJy.

The subarcsecond angular resolution observations at 3.6~cm failed to detect 
emission associated with either the infrared sources in this region or
with HH~68b, within an upper limit of $\sim 0.10$~\mjy\ (at 4-$\sigma$ level).
Given the quite different angular resolution of Avila \et\ (2001) observations,
$\sim3''$, and our observations, the non-detection of HH~68b at subarcsecond
angular resolution suggest that the weak emission associated with this
HH object is either extended or time variable.

\begin{figure}
\includegraphics[width=\columnwidth]{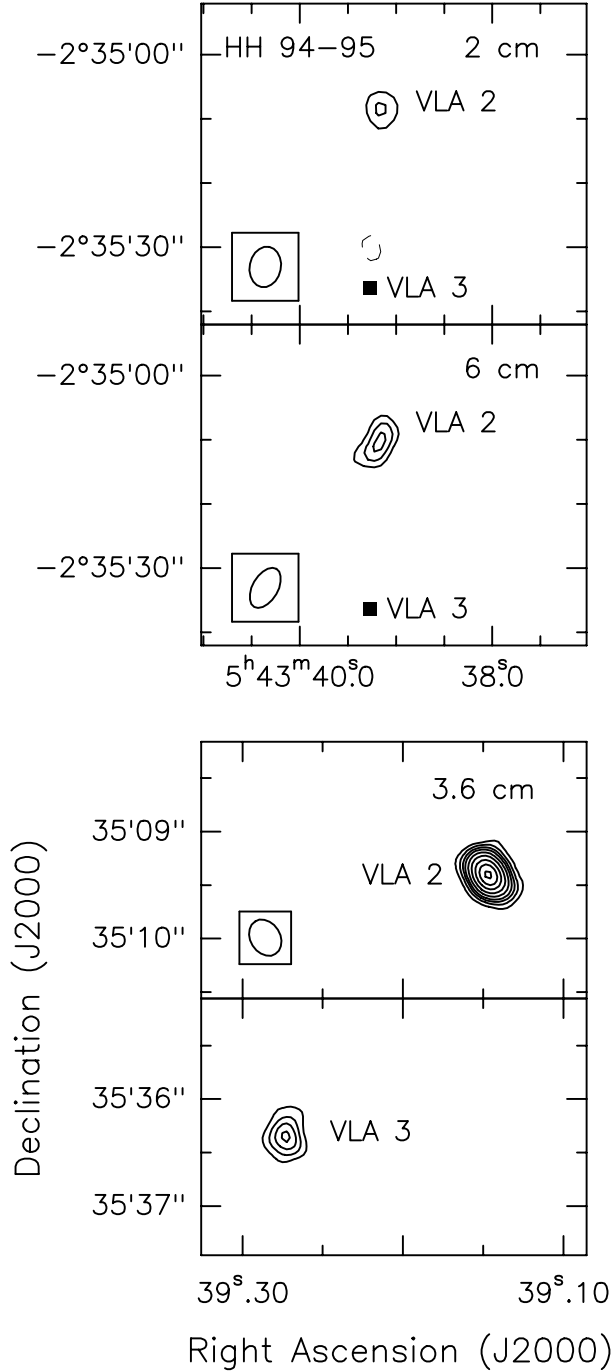}
\caption
{Composite of VLA maps in the HH~94-95 region: 
{\it Two top panels:} 2~cm and 6~cm matching--beam 
maps. Contours are $-$3, $-$2, 2, 3, 5, and 7 times the rms of the map, 61 
and 36~$\mu$Jy~beam$^{-1}$ at 2 and 6~cm, respectively.
{\it Two bottom panels:} the subarcsecond angular angular resolution maps 
at 3.6~cm around VLA sources VLA~2  and VLA~3.  Contours are $-$3, 3, 5, 7, 9,
12, 16, 21, 26, 31, 36 times the rms of the map, 9~$\mu$Jy~beam$^{-1}$.  
The half power contour of the synthesized beams are shown in the bottom left 
corner of each map.
\label{hh94} 
}
\end{figure}

\subsection{\hhn}

HH~94 and 95 are two Herbig-Haro objects (also known as Re~56 and 57),
separated by 8$'$, located in the L1630 molecular cloud in Orion. The detection
of these two HH objects was first reported by Reipurth (1985).  HH~95 has a
clear bow  shock morphology facing away from HH~94, which suggests that both
HH  objects could be driven by the same source (Reipurth 1989). Curiel \et\
(1989b) detected a faint radio continuum source located near the geometric
center  of this HH system, and proposed it as the powering source of both HH
objects. Nevertheless, sensitive mm observations failed to show neither dust
emission at this position nor molecular line emission associated with this
object  (Dent, Matthews \& Ward-Thompson 1998).

The matching--beam and the subarcsecond (3.6~cm) radio continuum maps of this 
region are shown in Figure~\ref{hh94}.  The 3.6~cm observations were carried 
out in two epochs, 1990 and 1997.  Source VLA~2, the suggested driving source 
(Curiel \et\ 1989b), does not show any variation between these  two
observations and appears unresolved, with an upper limit of the emitting size
of $\sim0\farcs1$ or 50~AU.  The flux density measured at the three wavelengths
observed is consistent with a flat spectrum, $\alpha=0.0\pm0.3$.  At about
27$''$ south of VLA~2, we have detected a new weak source from our 1997 
3.6~cm  observations, with a peak intensity of $0.08\pm0.01$~\mjy\ and a total 
flux density of $0.12\pm0.02$ mJy.  The upper limit for this source from the 
Curiel \et\ (1989b) observations, $\sim 0.09$~\mjy\ (5-$\sigma$), is consistent
with our observations.  Assuming that this source is not variable, the flux
measured at 3.6~cm together with the 2 and 6~cm upper limits imply a spectral
index between $-0.4$ and 1.3. 

Given the properties of the emission of sources VLA~2 and VLA~3 and their
location close to the geometrical center of the HH~94-95, these two objects
are  possible powering source candidates.  However, Dent \et\ (1998) did not
detected dust emission towards VLA~2, although their observations did not
include VLA~3.  A YSO powering an HH system is expected to have detectable dust
emission (see Reipurth et\ 1993). Further observations are needed to determine
if VLA~3 has dust emission associated, and therefore is the powering
source of the HH object.

\subsection{NGC~2264D}

\begin{figure*}[t]
\vspace{30mm}
\includegraphics[angle=-90,scale=0.88]{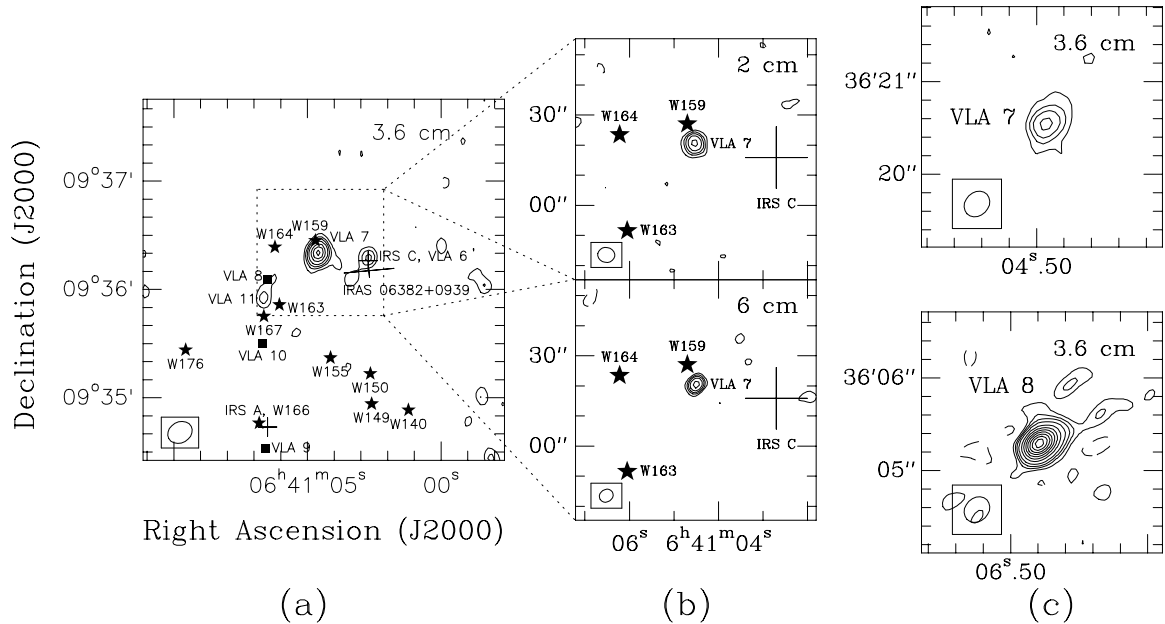}
\vspace{-35mm}
\caption
{Composite of VLA maps in the NGC~2264D region: 
{\it (a)} The low angular resolution maps at 3.6~cm, 
{\it (b)} the 2~cm (bottom panel) and 6~cm (top panel) matching--beam maps 
around VLA~7,
{\it (b)} Subarcsecond angular resolution map at 3.6~cm around VLA sources 
VLA~7 (top panel), VLA~8 (bottom panel). Contours are $-$3, 3, 5, 7, 10, 15, 
20, 25, 30, 35, 40 and 45 times the rms noise of the maps.  The rms of the high
and low angular resolution maps at 3.6~cm are 42 and 24~$\mu$Jy~beam$^{-1}$,
respectively. The rms for the 2 and 6~cm are 59, 41~$\mu$Jy~beam$^{-1}$, 
respectively.  Position of stars and infrared sources are shown as stars and 
crosses respectively (L\'opez-Molina \et\ 1990, Mendoza \et\ 1990).  The half 
power contour of the synthesized beams are shown in the bottom left corner of 
each map.
\label{n2264d} 
}
\end{figure*}

NGC~2264D is the most massive bipolar molecular outflow located in the
Monoceros OB1 molecular cloud (Margulis \& Lada 1986).  Not  far from the
geometrical center of this molecular outflow there is an  IRAS point source,
IRAS~06382+0939, with an infrared luminosity of  $\sim$ 550 L$_{\odot}$
(Castelaz \& Grasdalen 1988).  Far-infrared airborne observations show that the
emission extends over several arc minutes (Cohen, Harvey \& Schwartz 1985).  
Mendoza \et\ (1990) detected a cluster of near-infrared  objects.  Four of them
are spatially coincident with the molecular outflow.  Two of these sources,
IRS~A and IRS~C appear to be close to the geometrical center of the molecular
outflow.  IRS~C is associated with IRAS~06382+0939 and exhibits X-ray emission
(Nakano \et\ 2000). IRS~A is coincident with W166, a Herbig Be/Ae star (Neri,
Chavarr\'{\i}a-K. \& de Lara 1993). Previous VLA observations (Mendoza \et\
1990) showed only one radio continuum source close to the center of the
molecular outflow, $\sim$30$''$ east from IRAS~06382+0939  (IRS~C).  To the
north of the molecular outflow there are three HH objects (HH~125, HH~225 and
HH~226) which are likely part of the same optical outflow system (Walsh, Ogura,
\& Reipurth 1992). Two infrared objects, IRAS~06382+0939 and IRAS~06382+0945,
have been suggested as the driving source of this optical outflow (Cohen,
Harvey \&  Schwartz 1985; Walsh, Ogura, \& Reipurth 1992).

\begin{table}
\scriptsize
\caption[]{Sources close to the center of the molecular outflow in NGC~2264D.}
\label{tn2264d}
\begin{tabular}{ccccc}
\hline
\multicolumn{1}{c}{} &
\multicolumn{4}{c}{Flux density (mJy)\tablenotemark{a}}
\\
\multicolumn{1}{c}{} &
\multicolumn{4}{c}{\hrulefill}
\\
\multicolumn{1}{c}{} &
\multicolumn{1}{c}{2-cm} &
\multicolumn{1}{c}{3.6-cm} &
\multicolumn{1}{c}{3.6-cm} &
\multicolumn{1}{c}{6-cm}
\\
\multicolumn{1}{c}{Source} &
\multicolumn{1}{c}{($\sim 5''$)} &
\multicolumn{1}{c}{($\sim 0\farcs3$)} &
\multicolumn{1}{c}{($\sim 10''$)} &
\multicolumn{1}{c}{($\sim 5''$)}
\\
\hline
6&       $\la0.25$&    $\la 0.09$& $0.32\pm0.04$&     $\la0.16$ \\
7&   $1.12\pm0.08$& $0.62\pm0.08$& $1.17\pm0.07$& $0.72\pm0.04$ \\
8&       $\la0.25$& $1.52\pm0.06$&    $\sim0.13$&     $\la0.16$ \\
9&\tablenotemark{b}&$0.21\pm0.06$&     $\la0.24$&     $\la0.18$ \\
10&      $\la0.32$& $0.20\pm0.05$&     $\la0.19$&     $\la0.17$ \\
11&      $\la0.26$&     $\la0.10$& $0.26\pm0.04$&     $\la0.16$ \\
\hline
\end{tabular}
 $^a$ Upper limits are at 4-$\sigma$ level, taking into account the
primary beam response. \\
 $^b$ Outside of the HPFW of the primary beam. \\
\end{table}

From the different observations that we have carried out, we have detected
6 sources towards the center of the molecular outflow
(see Table~\ref{tn2264d} and Figure~\ref{n2264d}): 
\\
{\em Source VLA~6} was detected only in the 3.6~cm low angular resolution map. 
This source is located $1\farcs7$ from IRS~C.  Given the position uncertainties
of the infrared position (Castelaz \& Grasdalen 1988), these two sources are
probably related.
The upper limits obtained from the other 3 maps (6, 3.6 and 2~cm) suggest
that the emission of this source could be variable, or that the source has
a nearly flat spectral index and it is resolved out in the high resolution 
3.6~cm observations.
\\
{\em Source VLA~7} is located $\sim 7''$ south of the star W159
(L\'opez-Molina, Neri \& Chavarr\'{\i}a-K. 1990).  The position accuracy of the
3.6~cm  subarcsecond map ($<0\farcs1$) and of the optical image ($<1''$: 
L\'opez-Molina \et\ 1990) suggest that they are probably not associated. No
infrared object is associated with this source.  The subarcsecond resolution 
map reveals that source VLA~7 is partially resolved in the north-south
direction, with a deconvolved size of $0\farcs46\pm0\farcs02 \times
0\farcs27\pm0\farcs02$ and a position angle of  $PA=162\arcdeg\pm5\arcdeg$.
From the matching--beam maps we derive a spectral index of $\alpha = 0.4 \pm
0.2$, consistent with it being partially thick thermal emission. The two flux
densities measured at 3.6~cm differ by almost a factor 2, which could be the
result of the source being partially resolved out in the subarcsecond
resolution map or, to variability of this source.  Within the uncertainties,
the flux density of the lower angular resolution 3.6~cm image is roughly in
agreement with the derived matching-beam spectral index.
\\
{\em Sources VLA~8 and VLA~11} were only detected at 3.6~cm (the former in both
high and lower angular resolution maps, and the latter only in the lower
resolution map).  They do not have an optical or infrared counterpart.
Source VLA~8 increased its 3.6~cm flux by a factor of 10 between 1992 and 1995. 
The emission of this source is partially resolved, with a deconvolved size of 
$0\farcs21\pm0\farcs01\times0\farcs07\pm0\farcs01$, $PA=128\arcdeg\pm3\arcdeg$.
However, this elongation and orientation is likely due to beam smearing (the
source is located 32$''$ south-east from the phase center).
\\
{\em Sources VLA~9 and VLA~10} were only marginally detected in the 3.6~cm
subarcsecond resolution maps. They do not have any counterpart associated
to them, although they appear located within a dense core 
(Wolf-Chase, Walker, \& Lada 1995). 

The radio continuum source VLA~7 is a good candidate for the exciting source of
the HH~125/225/226 outflow system: it is well aligned with the three HH objects
and the emission is partially resolved in the direction of the HH outflow and
it has a positive spectral index (derived from the matching--beam observation)
which probably indicates free-free emission.  Based on its closest location to
the  geometrical center of the molecular outflow (Mendoza \et\ 1990), IRS~A and
W166 are also good candidates for powering the molecular outflow. However,  we
did not detect radio continuum emission associated with these sources.
Alternatively, radio sources VLA~9 or VLA~10 could also be candidates for the
driving source of the molecular outflow, since they appear to be embedded in a
dense core (Wolf-Chase, Walker, \& Lada 1995), but further observations are
required to confirm the detections.

\subsection{L1681B: YLW~16A}

\begin{figure}
\includegraphics[width=\columnwidth]{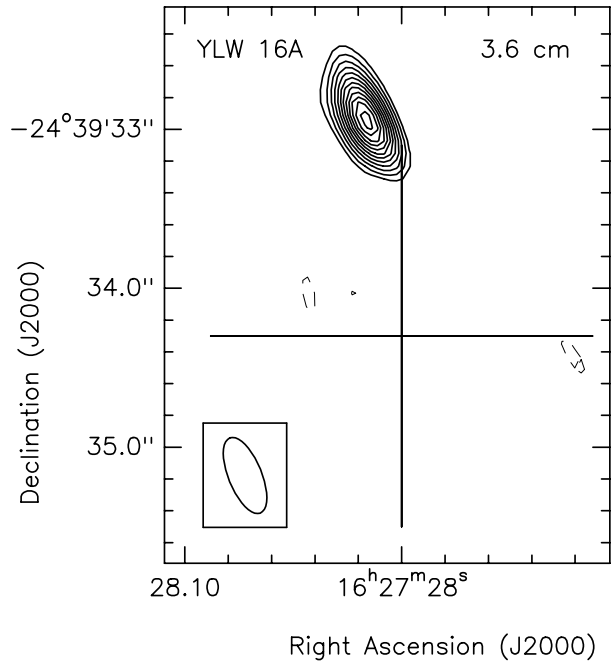}
\caption
{Subarcsecond angular resolution map at 3.6~cm of YLW~16A.  Contours are -3, 3,
5, 7, 9, ... 27 times the rms noise  of the map, 17~$\mu$Jy~beam$^{-1}$.  The
half power contour of the synthesized  beam is shown in the bottom left
corner.  The cross marks the 1-$\sigma$ position uncertainty of the near
infrared YSO (Barsony et al. 1997).
\label{ylw} 
}
\end{figure}

YLW~16A (IRS~44) is a Class I source with total  luminosity of 13~\lo, located
in the L1681B molecular  cloud within the $\rho$ Ophiuchi molecular cloud
complex (Wilking, Lada \&  Young 1989; Andr\'e \& Montmerle 1994).  It has
associated X-ray emission (Casanova \et\ 1995; Kamata \et\ 1997; Grosso 2001),
a compact molecular outflow (Bontemps \et\ 1996; Sekimoto \et\ 1997) and water
masers (Wilking \& Claussen 1987).  Observations at moderate angular resolution
($\sim 10''$) show there is radio continuum emission associated with YLW~16A
(Andr\'e, Montmerle \& Feigelson 1987; Leous \et\ 1991).  

Figure~\ref{ylw} shows the 3.6~cm subarcsecond resolution maps of YLW~16A. 
Within the primary beam we also detected YLW~15, which has been reported by 
Girart, \ro\ \& Curiel (2000).  YLW~16A appears marginally resolved, with a
deconvolved size of  $0\farcs27\pm0\farcs02\times0\farcs13\pm0\farcs02$ and a
position angle of $47\arcdeg \pm7\arcdeg$.  The molecular outflow is close to a
pole-on configuration (Sekimoto \et\ 1997), so it is difficult to define the
axis of symmetry of the molecular outflow in the plane of the sky and compare
it with the radio emission elongation direction. In order to compare the VLA
flux density of YLW~16A with  previous, lower angular resolution measurements,
maps were done by applying a  Gaussian taper to the visibilities.  The flux
measured is $0.78\pm0.06$~\mjy.  This value is consistent with the previous VLA
observations (Andr\'e \et\ 1987; Leous  \et\ 1991; Andr\'e \et\ 1992).  From
our fluxes and the previous ones obtained at other frequencies we estimate
an spectral index of $-0.09\pm0.11$ for YLW~16A, which is well consistent
with optically thin free-free emission.

\subsection{L778}

\begin{figure}
\includegraphics[width=\columnwidth]{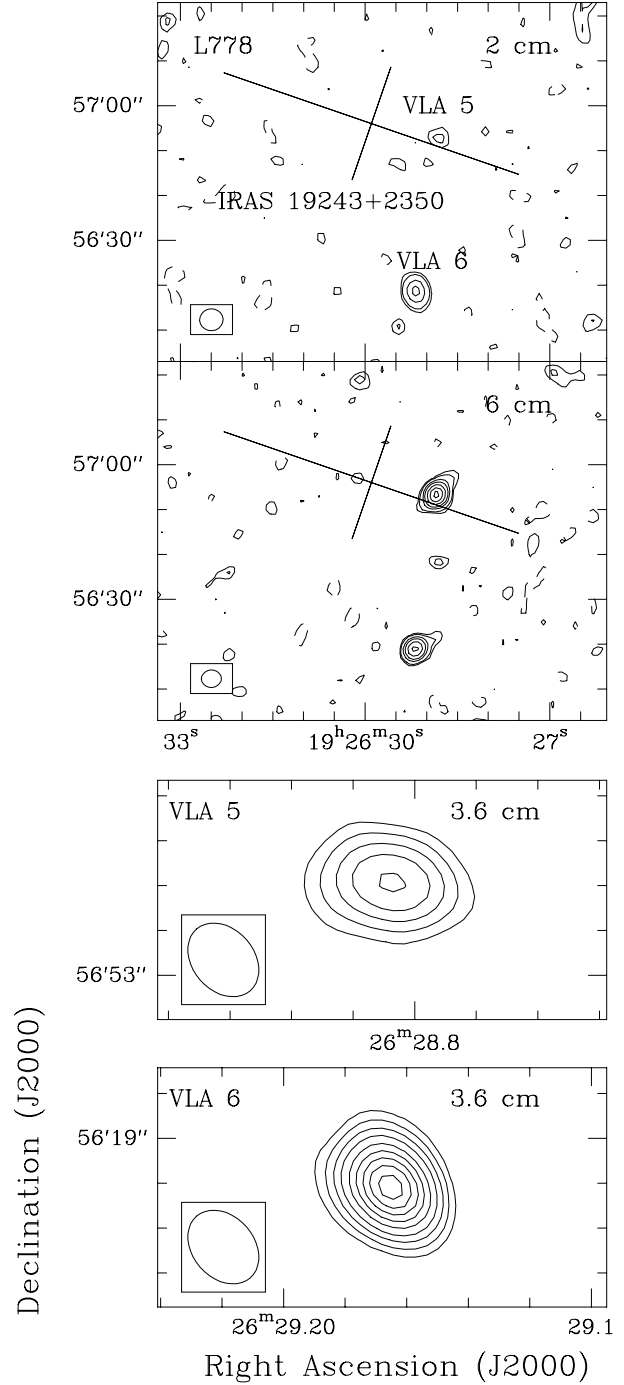}
\caption
{Composite of VLA maps in the L778 region. {\it Two top panels:} 2~cm and 6~cm
matching--beam maps. Contours are -2, 2, 3, 6, 9, 14, 20 and 26 times the rms
of the  maps: 57 and 33~$\mu$Jy~beam$^{-1}$ at 2 and 6~cm, respectively. {\it
Two bottom panels:} Subarcsecond angular resolution map at 3.6~cm of VLA 
sources VLA~6 and VLA~5.  Contours are  5, 9, 14, 20, 28, 36, 44, 52 and 60
times  the rms noise of the maps, 15~$\mu$Jy~beam$^{-1}$.  The half power
contour of  the synthesized beams are shown in the bottom left corner of each
map.  
\label{l778} 
}
\end{figure}

L778 is a dark cloud located in Vulpecula at a distance of about 250 pc
(Beichman \et\ 1986).  It has an ammonia dense core and a complex CO outflow, 
with two distinct blue lobes and one red lobe (Myers \et\ 1988).  The infrared
source IRAS~19243+2352 is well centered on the core (Myers \et\ 1988). \ro\
\et\ (1989) detected three radio continuum sources in this region, one of them
within the ellipsoid error of the infrared source IRAS~19243+2350, located
$\sim1\farcm2$ to the south of IRAS~19243+2352 and the ammonia core.
IRAS~19243+2350 appears closer to the geometrical center of the CO molecular
outflow, so \ro\ \et\ (1989) suggested that this radio continuum source is also
a viable candidate for driving the molecular outflow.

We detected a total of 7 sources in the 6~cm field, including the three radio 
sources previously reported by \ro\ \et\ (1989).  Sources VLA~5 and VLA~6,
(see  Table~\ref{tmatch} and \ref{thigh}) lie close to the center of  the
molecular outflow. Figure~\ref{l778} shows the 2 and 6~cm matching--beam  and
the 3.6~cm subarcsecond resolution maps of these two sources.  Source VLA~5,
apparently associated with IRAS~19423+2350, is just marginally detected  at
2~cm and exhibits a negative spectral index (see Table~\ref{tsize}). Its
emission is clearly resolved in the 3.6~cm, with an elongated morphology.  Its
deconvolved size is
$\theta_{ds}=0\farcs34\pm0\farcs01\times0\farcs09\pm0\farcs02$
($85\times23$~AU  in projection assuming a distance of 250~pc) with a position
angle of  $PA=97\arcdeg\pm2\arcdeg$.  Source VLA~6 has a positive spectral
index (see  Table~\ref{tsize}) and at high angular resolution appears
unresolved.  This source has no known counterparts at other wavelengths. 
Inspection of the 3.6~cm data (the one with the highest sensitivity) around
IRAS~19243+2352 and the center of the  ammonia core failed to detect any radio
continuum source stronger than $\sim0.1$~mJy (at 5-$\sigma$ level).

\begin{figure}[t]
\includegraphics[width=\columnwidth]{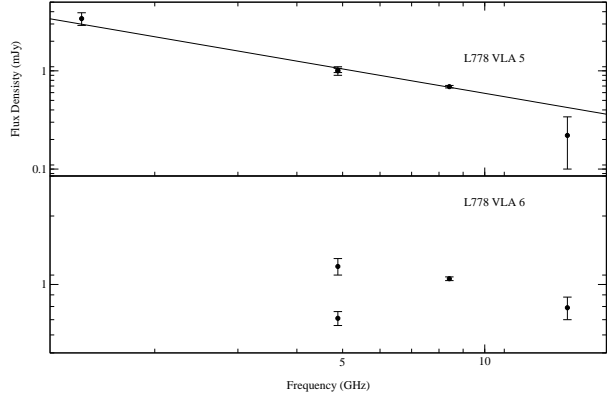}
\caption
{Centimeter spectrum of VLA 5 (top) and VLA 6 (bottom) in L778. For VLA 6 at
5~GHz the weaker point is from the flux measured in this paper, and the 
stronger point is the flux measured by  \ro\ \et\ (1989). The solid line shows
the best fir to the VLA 5 spectral index. 
\label{l778_si} 
}
\end{figure}

In order to better estimate the spectral indices of the detected sources and
check for variability we have taken into account previous measurements by \ro\ 
\et\ (1989) at 6~cm obtained with a $17''$ angular resolution, and the  20~cm
fluxes obtained from the NVSS survey (Condon \et\ 1998).   The different
measurements obtained for source VLA~5 do not suggest variability  and are in
agreement with a spectral index of $\alpha = -0.82\pm0.04$  (see
Table~\ref{tsize} and Fig.~\ref{l778_si}).  On the other hand, source VLA~6 is
clearly variable  (Fig~\ref{l778_si}), with a 40\% change in the flux (at
4-$\sigma$ level) in  a period of only 2 years.

Which is the powering source of the molecular L778 outflow? Source VLA~6 is not
far from the geometrical center of the southern blue and red lobes (see Myers
\et\ 1988), however its lack of infrared or submm counterpart and the
variability of the radio continuum emission suggest that this source is
probably not associated with the molecular outflow.  Source 5, associated with
IRAS~19243+2350, is elongated roughly in the same direction of the two lobes 
(blue and redshifted) of the CO molecular outflow. Yet, it has a clearly
negative spectral index. One possibility is that this source is a
pre-main-sequence star, since some post T Tauri (Class III) stars exhibit
non-thermal emission in the centimeter wavelength (e.g. White \et\ 1992). 
However, the spectral energy distribution of IRAS~19243+2350 (with fluxes of
$\la$0.33, 0.57, $\ga$0.40 and 9.10 Jy at 12, 25, 60 and 100 $\mu$m
respectively) are more consistent with a rather younger low mass YSO: a Class I
object.  Another possibility is that this YSO is powering a synchrotron radio
jet, as is the case of the H$_2$O maser source in W3(OH) (Reid \et\ 1995;
Wilner, Reid \& Menten 1998).  This source is a luminous young star that has a
synchrotron radio jet with an spectral index of $-0.6\pm0.1$ associated to it. 
This object is well modeled as a biconical synchrotron source tracing a fast
well collimated wind (Reid \et\ 1995; Wilner, Reid \& Menten 1998).  However,
since there is not known, up to date, a low-mass YSO powering a molecular
outflow with this type of emission, further observations are required in order
to check whether this mechanism can explain the radio properties of this source
or, alternatively, that the radio emission is not associated spatially with the
IRAS source.

\subsection{V645~Cyg}

\begin{figure}[t]
\includegraphics[width=\columnwidth]{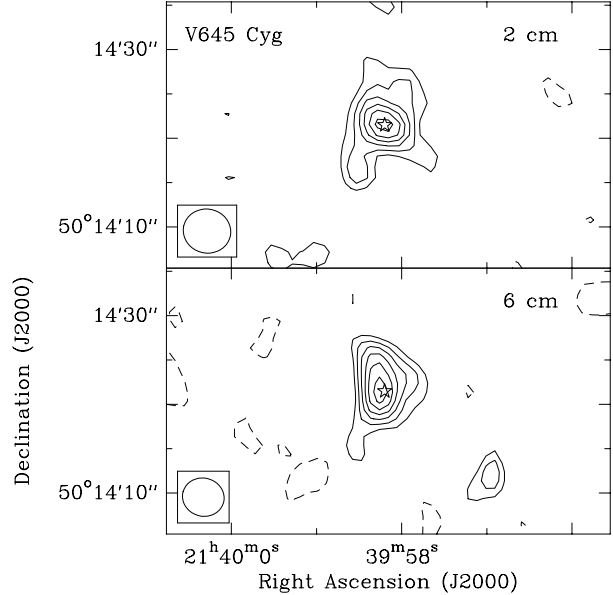}
\caption
{2~cm and 6~cm matching--beam  maps of \vcy. Contours are -2, 2, 3, 4, 5, 6 and
7 times the rms noise of  the maps, which is 66 (2~cm) and
32~$\mu$Jy~beam$^{-1}$ (6~cm).  The half power contour of  the synthesized
beam  is shown in the bottom left corner.  The stars mark the position of the
Herbig Ae/Be star in \vcy.  
\label{v645} 
}
\end{figure}

\vcy\ is an Herbig Ae/Be object with a spectral type between A0 and A5 and a
bolometric luminosity of $4\times10^4 [D/3.5 {\rm Kpc}]^2$~\lo, associated with
an optical extended nebulosity (\eg\ Cohen, 1977; Goodrich 1986; Natta \et\
1993).  Two different distances have been used in the bibliography for this
object: 6 and 3.5~Kpc.  Nevertheless, kinematic and extinction analysis suggest
that 3.5~Kpc is the  most likely value (Goodrich 1986; Schulz \et\ 1989) and
this is  the value adopted in this paper.
The star is surrounded by a high density molecular core of 
$\sim 80$\mo\ (\eg\ Torrelles \et\ 1989; Natta \et\  1993).  H$_2$O and an
unusual OH maser emission is associated with this source (Lada \et\ 1981;
Morris \& Kaz\`es 1982).  \vcy\  powers a low velocity bipolar molecular
outflow with a modest degree of collimation in the north-south direction (\eg\
Verdes-Montenegro \et\ 1991).  The high luminosity of the source suggests that
the radiation field from the star could be driving the molecular  outflow
(Verdes-Montenegro \et\ 1991). 

After several unsuccessful attempts (\eg\ Kwok 1981; \ro\ \&
Cant\'o 1983), the detection of the radio source associated with
\vcy\ was first reported by Curiel \et\ (1989a) from 6~cm
VLA observations.  They found that the radio source was slightly resolved at
an angular resolution of $\sim 10''$.  Skinner, Brown \& Stewart
(1993) confirmed that the radio continuum emission of \vcy\ is 
extended, especially in the north-south direction
($\sim 8\farcs7 \times 6\farcs6$) and found that the spectral index
is almost flat ($\alpha = -0.2\pm0.3$).

The matching--beam maps (see Figure~\ref{v645}) show that the emission is 
partially resolved, especially at 6~cm (see Table~\ref{tsize}), and roughly 
elongated in the North-South direction.  A Gaussian fit of the emission at 
6~cm gives a deconvolved size of 
$6\farcs5\pm0\farcs4 \times 3\farcs6\pm0\farcs4$ and a position angle of 
$PA=7\arcdeg\pm6\arcdeg$. From the matching--beam total fluxes, the 
derived spectral index is $\alpha = 0.5\pm0.2$ (Table~\ref{tsize}), which is 
significantly higher than the one obtained by Skinner \et\ (1993).

\begin{table}
\scriptsize
\caption[]{Observations of V645~Cyg at different epochs}
\label{tv645}
\begin{tabular}{lcccc}
\hline
\multicolumn{1}{c}{} &
\multicolumn{1}{c}{Angular} &
\multicolumn{2}{c}{Flux density (mJy)} &
\multicolumn{1}{c}{}
\\
\multicolumn{1}{c}{Epoch} &
\multicolumn{1}{c}{Resolution} &
\multicolumn{1}{c}{3.6-cm} &
\multicolumn{1}{c}{6-cm} &
\multicolumn{1}{c}{Ref.}
\\
\hline
Mar-1980 & $0\farcs5$ &              -&      $\la0.5$ & (1) \\
Oct-1981 &      $5''$ &              -&      $\la1.0$ & (2) \\
Oct-1984 &     $15''$ &              -& $1.02\pm0.04$ & (3)\tablenotemark{a} \\
Jul-1989 &      $5''$ &              -& $0.57\pm0.04$ & (4) \\
Feb-1990 & $0\farcs3$ &     $\la0.11$ &             - & (5) \\
Feb-1991 &      $7''$ & $0.76\pm0.15$ &             - & (5) \\
Feb-1991 &     $14''$ &             - & $0.87\pm0.21$ & (5) \\
Jun-1991 &     $14''$ & $0.70\pm0.21$ &             - & (5) \\
Jan-1994 &     $11''$ & $0.99\pm0.04$ &             - & (4) \\
Jan-1996 & $1\farcs4$ & $0.55\pm0.18$ &             - & (6) \\
\hline
\\
\end{tabular}
References: (1) Kwok 1981, (2) \ro\ \& Cant\'o 1983, (3) Curiel \et\
1989a, (4) this paper, (5) Skinner \et\ 1993, (6) Di Francesco \et\ 1997. \\
 $^a$ We re-made the map using natural weighting. \\
\end{table}

\begin{figure}[t]
\includegraphics[width=\columnwidth]{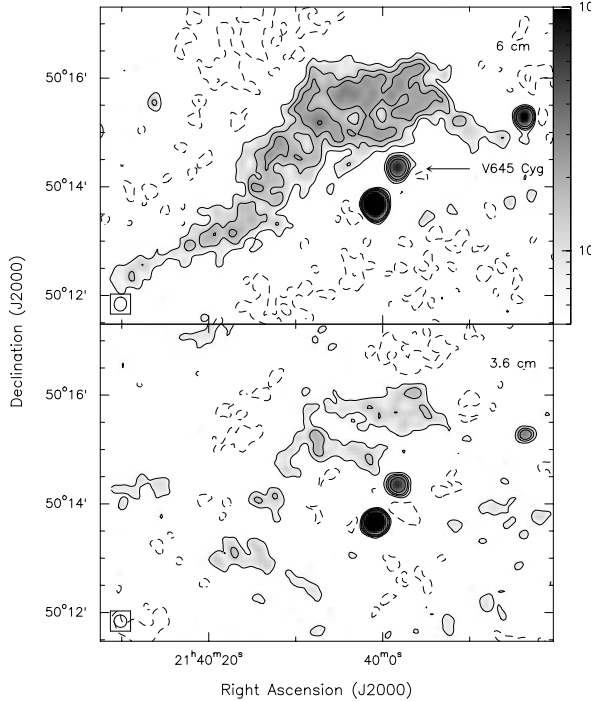}
\caption
{Cleaned, natural-weight D configuration VLA maps of the \vcy\ region at 3.6~cm
(bottom panel) and 6~cm (top panel).  Contours are -4, -2, 2, 4, 6, 10, 
15 and 20  times the rms noise of the maps, 34 and 32~$\mu$Jy~beam$^{-1}$ 
respectively.  The half power contour of the synthesized beam is shown in the 
bottom left corner.  
\label{v645_6cm} 
}
\end{figure}

We carried out additional VLA observations at 3.6~cm in D configuration which
provided an angular resolution of $\sim 11''$.  The emission is also partially
resolved with a deconvolved size of 
$8\farcs9\pm0\farcs2\times6\farcs0\pm0\farcs5$ and $PA=39\arcdeg \pm 5\arcdeg$,
which is very similar to that obtained by Skinner \et\ (1993) and Curiel \et\
(1989a) at 6~cm, but slightly larger than the value obtained from the
matching--beam 6~cm map.  The 3.6~cm flux density,  $0.99\pm0.04$~mJy, is
inconsistent with the matching--beam flux densities.  In  Table~\ref{tv645} we
list the different fluxes measured with the VLA at 3.6~cm  and 6~cm, the
observing date and the angular resolution of the observations.  The data on
this table seems to suggest that the flux density decreases with increasing
angular resolutions, especially for angular resolution of $\la 5''$.  In order
to check if the angular resolution is really affecting the  matching--beam flux,
we assume that the radio emission of \vcy\ is a Gaussian  with a full width at
half maximum of the peak emission of $\sim 9''$ (the major axis of the lower
angular resolution measurements).  The half-power radius  in the visibility
domain is:
\begin{equation}
r_{(u,v)} \, = \, 91.02 \, 
\left ( \frac{\theta_{\rm FWHM}}{''} \right )^{-1} \, 
{\rm k}\lambda 
\end{equation}
and the fraction of the total flux density observed for a visibility coverage 
with an inner radius, $r_{\rm min}$, is:
\begin{equation}
\frac{S_{observed}}{S_{total}} \, = \, 
e^{
- 8.4\times10^{-5} \,
(r_{\rm min}/{\rm k}\lambda)^2 \,
(\theta_{\rm FWHM}/'')^2 
}
\end{equation}
The matching--beam observations have a $(u,v)$ coverage radius from $\sim $1 to 
59~\kl\ at 6~cm, but it is for a radius $\ga 2$~\kl\ that the visibility 
distribution becomes roughly homogeneous in the $(u,v)$ plane.  Therefore,
using $r_{\rm min} = 2$~\kl, the flux detected should be 97\% of the total flux 
density at 6~cm.  Even for a source of $18''$ we would have detected 90\% of 
the total flux.  In addition, we made 6~cm matching-beam maps 
by applying a Gaussian taper of 25~\kl\ in the $(u,v)$ plane, which provided a 
synthesized beam of $\sim 8''$.  The new estimated flux densities do not 
show any significant change at 1-$\sigma$ level (0.04~\mjy).  Therefore, 
we conclude that the fluxes measured from the matching--beam observations are 
not underestimated. Thus, the differences observed in the flux density
measured at 6~cm (see Table~\ref{tv645}), especially between the
Oct-1984 and Jul-1989 observations (about $0.45\pm0.06$~mJy), are
likely to be due to time variability. We speculate that the variations
in flux density and size observed in this source could be the result of
episodic ejection of material by the powering source.  Another possibility is
that the ionized gas associated with this source can recombine in timescales of
a few years, which implies electron densities in excess of 
$3\times10^4$~cm$^{-3}$.  Further observations will be needed to test these
possibility.

\begin{figure}[t]
\includegraphics[width=\columnwidth]{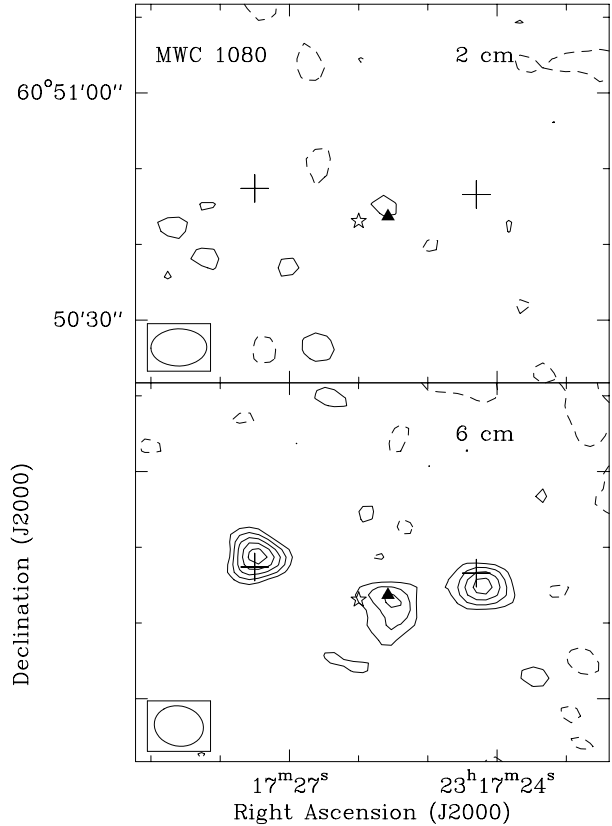}
\caption
{2~cm (top panel) and 6~cm (bottom panel) matching--beam maps of \mwc.  
Contours are -2, 2, 3, 4, 5, 6 and 7 times the rms noise of the maps,  
80 and 35~$\mu$Jy~beam$^{-1}$ at 2 and 6~cm, respectively.  The half power 
contour of the synthesized beam is shown in the bottom left corner.  The
filled triangle marks the position of the 3.6~cm radio source detected by
Skinner, Brown \& Linsky (1990).  The star marks the position of the
Herbig~Ae/Be star in \mwc\ (Hillenbrand \et\ 1992).  The two crosses mark the
position of dust emission peaks (Fuente \et\ 1998).
\label{m1080} 
}
\end{figure}

{\it Other objects in the field:}

From the VLA D configuration 6~cm map, we detect an extended source close to
V645Cyg with a size of $\sim$$6'\times2'$ (see Figure~\ref{v645_6cm}) and a
total flux density of $31\pm1$~mJy.  The position of this source coincides with
the radio object BWE~2138+5001 detected at 6~cm by  Becker, White \& Edwards
(1991) from low angular resolution ($\sim 3\farcm5$), single dish observations.
Using data from the NVSS (NRAO VLA Sky Survey) survey (Condon, \et\ 1998), we
obtain that BWE~2138+5001 has a flux density of $34\pm3$ at 20~cm. The
estimated spectral index between 6 and 20~cm ($\alpha=-0.1\pm0.1$) for this
extended object appears to be consistent with this source being an optically
thin HII region.

\subsection{MWC~1080}

\mwc\ is another Herbig Ae/Be star with an A1 spectral type and a bolometric
luminosity of $\sim 8\times10^3$~\lo\, associated with an optical nebulosity
(\eg\ Yoshida \et\ 1992; Hillenbrand \et\ 1992  and references therein) and
located at a distance of about 2.5~Kpc.  \mwc\ is, in fact, a triple system
(Pirzkal, Spillar  \& Dyck 1997; Leinert, Richichi \& Haas 1997)  that has
faint X-ray emission (Zinnecker \& Preibisch 1994).   \mwc\ shows several
signposts of strong mass loss: strong P~Cygni profiles in  the Balmer lines
(\eg\ Herbig 1960; Yoshida \et\  1992), a powerful molecular outflow (\eg\ 
Cant\'o \et\ 1984; Levreault 1988), a poorly  collimated but extremely fast
(v$_{\rm H_{\alpha}}$ up to 1100~\kms) HH  bipolar outflow (\eg\ Poetzel, Mundt
\& Ray 1992). The  molecular cloud associated with \mwc\ has several hundreds
of solar masses  (\eg\ Yoshida \et\ 1991), while, from dust emission 
observations, the circumstellar material around \mwc\ has only a few solar 
masses (Mannings 1994; Fuente \et\ 1998).  

Curiel \et\ (1989a) detected, from $10''$ resolution observations at 6~cm, a
radio source displaced about 13$''$ from the star, surrounded by weak extended
emission that roughly engulfs the star as well as its associated optical
nebulosity.  Higher angular resolution observations ($\sim 1''$) by Skinner,
Brown \& Linsky (1990) detected at 3.6~cm a weak source located at about 2$''$
from the optical source.  

At 6~cm we detect 3 sources close to \mwc\ (see Figure~\ref{m1080}).  These
sources coincide in position with the three relative maxima of the extended
radio emission detected at lower angular resolution by Curiel \et\ (1989a, see
Fig.2 of their paper).  Curiel \et\ (1989a) estimated a total flux of $\sim
2$~mJy for the extended radio emission, which is significantly higher than the
$0.53\pm0.08$~mJy flux density measured in our 6~cm maps. This is probably
because we are resolving out the weak extended component detected by Curiel
\et\ (1989a). VLA~4 coincides with the position given by Skinner \et\ (1990),
but it is located about $3\farcs5$ west of \mwc, so its direct association with
the Herbig Ae/Be star is unclear. However, the presence of a strong HH outflow
in the east-west direction associated with \mwc\ (Poetzel, Mundt \& Ray 1992)
suggests that the emission could be related with the optical outflow.   None of
the three 6~cm source were detected at 2~cm up to a $3-\sigma$ level of
0.24~\mjy.  This implies that they have an spectral index smaller than 0.6.
Interestingly, two relative intensity peaks of the dust continuum emission, at
1.3~mm, mapped by Fuente \et\ (1998) seem to coincide in position with the two
strongest radio components associated with \mwc\ (Fig.~\ref{m1080}).  This
suggests that these two radio continuum sources may trace embedded, very young
stellar objects.

\section{Summary and Conclusions}

Table~\ref{tsize} shows the radio continuum candidates to power the molecular
and/or HH outflows in the regions observed with the VLA.  Of the observed
regions, HH~68-69 is the only one that is not shown in this table because of
the lack of radio continuum emission from our observations. The main results
from our observations are:

(1) We detected three radio sources, whose properties are consistent with being
thermal radio jets: L1489, \ngc\ VLA~7 and YLW~16A. The length of these radio
jets are $\sim$40~AU for L1489 and YLW~16A and $\sim$370~AU for \ngc\ VLA~7,
which are typical values for this type of objects (Anglada 1996). These three
radio jets are associated with HH and molecular outflows. L1489
coincides with a compact molecular outflow (Hogerheijde \et\ 1998) and with
HH~360 (G\'omez \et\ 1997). \ngc\ VLA~7 is associated with the system
HH~125/225/226 (Walsh, Ogura, \& Reipurth 1992).  For these two regions, the
thermal radio jet is well aligned with the molecular or HH outflow.  The
thermal radio jet in YLW~16A is associated with a pole-on molecular outflow
(Sekimoto \et\ 1997).

(2) We detected a non-thermal radio jet in L778 that appears to be
associated with a Class I infrared source, IRAS~19243+2350.
This radio jet is elongated in the direction
of a pair of red and blue high velocity CO lobes (Myers \et\ 1988) centered
roughly in IRAS~19243+2350.  Thus, we suggest the first tentative detection of
a  non-thermal radio jet associated with a low mass protostar. 

(3) Our observations could not find clear candidates for the
\hhn\ system and the molecular outflow in \ngc.
In both regions there are several radio sources that could
trace the powering source of the outflows, but the lack of known counterpart or
the incomplete information of their radio emission properties does not allow to
discriminate between them. 

(4) \vcy\ shows radio emission with striking properties: its emission is quite
extended, $\sim23000$~AU, but at the same time it is variable. There is
also an extended source that appears to be an optically thin HII region.

(5) There is a $\sim 3\farcs5$ offset between the optical source \mwc\ and
VLA~4, so it is not clear if VLA~4 is directly associated with the star.  There
are two other radio sources, VLA~3 and 5, that are associated with mm sources
and could be tracing protostars.

\acknowledgments
JMG acknowledges support by RED-2000 from the Generalitat de Catalunya and by
DGICYT grant PB98-0670 (Spain). JMG thanks the hospitality and support of the
Instituto de Astronom\'{\i}a-UNAM.
SC and LFR acknowledge support from DGAPA, UNAM and CONACyT, M\'exico. 
JC acknowledges support from CONACyT grants 34566-E and 36573-E. 



\begin{thebibliography}

\bibitem{}
 Andr\'e, P. \& Montmerle, T. 1994, \apj, 420, 837
\bibitem{}
 Andr\'e, P., Montmerle, T., \& Feigelson, E. D. 1987, AJ, 93, 1182
\bibitem{}
 Andr\'e, P. 1997, in {\em Herbig-Haro Flows and the Birth of the
 Low Mass Stars}, IAU Symposium  n.182, B. Reipurth,\& C. Bertout (eds.), 
 p. 483
\bibitem{} 
 Anglada, G. 1996, in {\em Radio Emission from Stars and the Sun},
 ASP Conf. Ser. Vol. 93, A. R. Taylor \& J. M. Paredes, 3
\bibitem{}
 Anglada, G., \ro, L. F., Cant\'o, J.,  Estalella, R., \& Torrelles, J. M. 
 1992, \apj, 395, 494
\bibitem{}
 Anglada, G., \ro, L.  F., Girart, J. M.,  Estalella, R., \& Torrelles, J. M. 
 1994, \apj, 420, L91
\bibitem{}
 Anglada, G., Villuendas, E., Estalella, R., Beltr\'an, M. T., \ro, L. F., 
 Torrelles, J. M., \& Curiel, S. 1998, AJ, 116, 2953
\bibitem{}
 Avila, R., \ro, L.~F., \& Curiel, S. 2001, \rmaa, 37, 201
\bibitem{} 
 Barsony, M., Kenyon, S.~J., Lada, E.~A., \& Teuben, P.~J.\ 1997, \apjs, 112, 
 109 
\bibitem{}
 Beichman, C. A., Myers, P. C., Emerson, J. P., Harris, S., Mathieu, R., 
 Benson, P. J., \& Jennings, R. E. 1986, \apj, 307, 337
\bibitem{}
 Becker, R. H., White, R. L., \& Edwards, A. L. 1991, ApJSS, 75, 1
\bibitem{}
 Beltr{\' a}n, M.~T., Estalella, R., Anglada, G., \ro, L.~F., \&
 Torrelles, J. 2001, \aj, 121, 1556
\bibitem{}
 Bontemps, S., Andr\'e, P., Tereby, S., \& Cabrit, S. 1996, A\&A, 858, 872
\bibitem{}
 Cabrit, S., \& Andr\'e, P. 1991, \apj, 379, L25
\bibitem{}
 Casanova, S., Montmerle, T., Feigelson, E. D., Andr\'e, P. 1995, \apj, 439, 
 752
\bibitem{}
 Cant\'o, J., \ro, L. F., Calvet, N., \& Levreault, R. M. 1984, \apj, 282, 631
\bibitem{}
 Castelaz, M. W., \& Grasdalen, G. 1988, \apj, 335, 150
\bibitem{}
 Condon, J. J. 1984, \apj, 287, 461
\bibitem{}
 Condon, J. J., Cotton, W. D., Greisen, E. W., Yin, Q. F., Perley, R. A., 
 Taylor, G. B., \& Broderick, J. J. 1998, AJ, 115, 1693
\bibitem{}
 Cohen, M., 1977, \apj, 215, 533
\bibitem{}
 Cohen, M. 1990, \apj, 354, 701
\bibitem{}
 Cohen, M., Harvey, P. M., \& Schwartz, R. D. 1985, \apj, 296, 633
\bibitem{}
 Curiel, S., Raymond, J.C., \ro, L.  F., Cant\'o, J., \& Moran, J. M. 1990, 
 \apj, 365, L85
\bibitem{a}
 Curiel, S., \ro, L. F., Cant\'o, J., Bohigas, J., Roth, M., \& Torrelles, 
 J. M. 1989a, Ap. Lett. Comm., 27, 299
\bibitem{b}
 Curiel, S., \ro, L. F., Cant\'o, J., \& Torrelles, J. M. 1989b, \rmaa, 17, 137
\bibitem{}
 Davis, C. J., \& Eisl\"{o}ffel, J. 1995, \apj, 443, L41
\bibitem{}
 Dent, W. R. F., Matthews, H. E.,  \& Ward-Thompson, D. 1998, MNRAS, 301, 1049
\bibitem{}
 Di Francesco, J., Evans II, N. J., Harvey, P. M., Mundy, L. G.,
 Guilloteau, S., \& Chandler, C. 1997, \apj, 482, 433
\bibitem{}
 Fuente, A., Mart\'{\i}n-Pintado, J., Bachiller, R., Neri, R., \& Palla, F.
 1998, A\&A, 334, 253
\bibitem{}
 Girart, J. M., \ro, L.~F., \& Curiel, S.\ 2000, ApJ, 544, L153
\bibitem{}
 G\'omez, J. F., Curiel, S., Torrelles, J. M., \ro, L. F., Anglada, G., \&
 Girart, J. M. 1994, \apj, 438, 749
\bibitem{}
 G\'omez, M., Whitney, B. A., Kenyon, S. J. 1997, AJ, 114, 1138
\bibitem{}
 Goodrich, R. W. 1986, \apj, 311, 882
\bibitem{}
 Grosso, N. 2001, A\&A, 370, L22
\bibitem{}
 Hartigan, P., Bally, J., Reipurth, B. \&  Morse, J. 2000, in {\em
 Protostars and Planets IV}, V. Mannings, A. Boss \& S. Russell  (eds.), 
 (University of Arizona Press), p. 841
\bibitem{}
 Herbig, G. H. 1960, ApJSS, 337
\bibitem{}
 Heyer, M. H., Ladd, E. F., Myers, P. C., \& Campbell, B. 1990, AJ, 99, 1585
\bibitem{}
 Hillenbrand, L. A., Strom, S. E., Vrba, F. J., \& Keene, J. 1992, \apj, 397,
 613 
\bibitem{}
 Hogerheijde, M. R., van Dishoeck, E. F., Blake, G. A., \& van
 Langevelde, H. J. 1998, \apj, 502, 315
\bibitem{}
 Hogerheijde, M. R., \& Sandell, G. 2000, \apj, 534, 880
\bibitem{Ket97}
 Kamata, Y., Koyama, K., Tsuboi, Y., Yamauchi, S. 1997, PASJ, 49, 461
\bibitem{}
 Kwok, S. 1981, PASP, 93, 361
\bibitem{}
 Lada, C. J., \& Lada, E. A. 1991, in {\em The formation and evolution of
 stars clusters}, p.3
\bibitem{}
 Lada, C. J., Blitz, L., Reid, M. J., Moran, J. M. 1981, \apj, 243, 769
\bibitem{}
 Leinert, C., Richichi, A., \& Haas, M. 1997, A\&A, 318, 472
\bibitem{}
 Leous, J. A., Feigelson, E. D., Andr\'e, P., \& Montmerle, T. 1991, \apj,
 379, 683
\bibitem{}
 Levreault, R. M. 1988, APJSS, 67, 283
\bibitem{}
 L\'opez-Molina, M. G., Neri, L., \& Chavarr\'{\i}a-K., C. 1990, \rmaa, 20, 113
\bibitem{}
 Lucas, P. W., Blundell, K. M., \& Roche, P. F. 2000, MNRAS, 318, 526
\bibitem{}
 Mannings, V. 1994, MNRAS, 271, 587
\bibitem{}
 Margulis, M., \& Lada, C. J. 1986, \apj, 309, L87
\bibitem{}
 Margulis, M., Lada, C. J., \& Snell, R. L. 1988, \apj, 333, 316
\bibitem{}
 Mendoza, E. E., \ro, L. F., Chavarria, C., \& Neri, L. 1990, MNRAS, 246, 518
\bibitem{}
 Miyawaki, R., \& Hayashi, M. 1992, PASJ, 44, 557
\bibitem{}
 Morris, M., \& Kaz\`es, I. 1981, A\&A, 111, 239
\bibitem{}
 Myers, P. C., Fuller, G. A., Mathieu, R. D., Beichman, C. A., Benson, P. J.,
 Schild, R. E., \& Emerson, J. P. 1987, \apj, 319,340
\bibitem{}
 Myers, P. C., Heyer, M., Snell, R. L., \& Goldsmith, P. F. 1988, \apj,
 324, 907
\bibitem{}
 Nakano, M., Yamauchi, S., Sugitani, K., \& Ogura, K. 2000, PASJ, 52, 437
\bibitem{}
 Natta, A., Palla, F., Butner, H.M., Evans II, N. J., \& Harvey, P. M.
 1993, \apj, 406, 674
\bibitem{}
 Neri, L. J., Chavarr\'{\i}a-K., C., \& de Lara, E. 1993, A\&ASS, 102, 201
\bibitem{}
 Ogura, K., \& Sugitani, K. 1998, PASA, 15, 91
\bibitem{}
 Ohashi, N., Hayashi, M., Kawabe, R., \& Ishiguro, M. 1996, \apj, 466, 317
\bibitem{}
 Padgett, D. L., Brandner, W., Stapelfeldt, K. R., Strom, S. E., 
 Tereby, S., Koerner, D. 1999, AJ, 117, 1490
\bibitem{}
 Pirzkal, N., Spillar, E. J.,  \& Dyck, H. M. 1997, \apj, 481, 392
\bibitem{}
 Poetzel, R., Mundt, R., \& Ray, T. P. 1992, A\&A, 262, 229
\bibitem{}
 Pravdo, S. H., \ro, L. F., Curiel, S., Cant\'o, J., Torrelles, J. M.,
 Becker, R. H., \& Sellgren, K.  1985, \apj, 293, L35
\bibitem{}
 Reid, M. J., Argon, A. L., Masson, C. R., Menten, K. M., \& Moran, J. M. 
 1995, \apj, 443, 238
\bibitem{}
 Reipurth, B. 1985, A\&AS, 61, 319
\bibitem{}
 Reipurth, B. 1989, A\&A, 220, 249
\bibitem{}
 Reipurth, B., \& Graham, J. A. 1988, A\&A, 202, 219
\bibitem{}
 Richer, J., Shepherd, D., Cabrit, S., Bachiller, R., \& Churchwell, E. 2000,
 in {\em Protostars and Planets IV}, V. Mannings, A. Boss \& S. Russell 
 (eds.), (University of Arizona Press), p. 867
\bibitem{}
 \ro, L. F. 1997, in {\it Herbig-Haro Flows and the Birth of the Low Mass
 Stars}, IAU Symposium  n.182, B. Reipurth,\& C. Bertout (eds.), p. 83
\bibitem{}
 \ro, L. F., \& Cant\'o, J. 1983, \rmaa, 8, 163
\bibitem{}
 \ro, L. F., Myers, P. C., Cruz-Gonz\'alez, I., \& Tereby, S.  1989, \apj,
 347, 461
\bibitem{}
 \ro, L. F., \& Reipurth, B. 1994, A\&A, 281, 882
\bibitem{}
 \ro, L. F., \& Reipurth, B. 1998, \rmaa, 34, 13
\bibitem{}
 \ro, L.~F.~et al.\ 1998, Nature, 395, 355
\bibitem{}
 Sekimoto, Y., Tatematsu, K., Umemoto, T., Koyama, K., Tsuboi, Y.,
 Hirano, N., Yamamoto, S. 1997, \apj, 489, L63
\bibitem{}
 Skinner, S. L., Brown, A., \& Linsky, J. L. 1990, ApJ, 357, L39
\bibitem{}
 Skinner, S. L., Brown, A., \& Stewart, R. T. 1993, ApJSS, 87, 217
\bibitem{}
 Schulz, A., Black, J. H., Lada, C. J., Ulich, B. L., Martin, R. N.,
 Snell, R. L., \& Erickson, N. J. 1989, \apj, 341, 288
\bibitem{}
 Stine, P. C., Feigelson, E. D., Andr\'e, P., \& Montmerle, T. 1988, AJ, 96, 
 1394
\bibitem{}
 Torrelles, J.~M., Verdes-Montenegro, L., Ho, P.~T.~P., \ro, L.~F., \&
 Canto, J.\ 1989, \apj, 346, 756. 
\bibitem{}
 Verdes-Montenegro, L., G\'omez, J. F., Torrelles, J. M.,
 Anglada, G., Estalella, R., \& L\'opez, R. 1991, A\&A, 244, 84
\bibitem{}
 Walker, C. K., Adams, F. C., \& Lada, C. J. 1990, \apj, 349, 515
\bibitem{}
 Walsh, J. R., Ogura, K., \& Reipurth, B. 1992, MNRAS, 257, 110
\bibitem{}
 White, S. M., Pallavicini, R., Kundu, M. R. 1992, A\&A, 257, 557
\bibitem{}
 Wilking, B. A., Lada, C. J., \& Young, E. T. 1989, \apj, 340, 823
\bibitem{}
 Wilking, B. A., \& Claussen, M. J. 1987, 320, L133
\bibitem{}
 Wilner, D. J., \& Lay, O. P. 2000, in {\it Protostars and Planets IV}, 
 V. Mannings, A. Boss \& S. Russell  (eds.), (University of Arizona Press), 
 p. 509
\bibitem{}
 Wilner, D. J., Reid, M. J., Menten, K. M. 1999, \apj, 513, 775
\bibitem{}
 Wolf-Chase, G. A., Walker, C. K., \& Lada, C. J. 1995, \apj, 442, 197
\bibitem{} 
 Wood, K., Smith, D., Whitney, B., Stassun, K., Kenyon, S. J., Wolff, M. J., 
 \& Bjorkman, K.S. 2001, \apj, 561, 299 
\bibitem{}
 Yoshida, S. Kogure, T., Nakano, M., Tatematsu, K., \& Wiramihardja, S. D. 
 1991, PASJ, 43, 362
\bibitem{}
 Yoshida, S. Kogure, T., Nakano, M., Tatematsu, K., \& Wiramihardja, S. D. 
 1992, PASJ, 44, 77
\bibitem{}
 Zhou, S., Wu, Y., Evans II, N. J., Fuller, G. A., \& Myers, P. C. \apj, 
 346, 168
\bibitem{}
 Zinnecker, H., Bastien, P., Arcoragi, J. P. \& Yorke, H. W. 1992, A\&A, 265, 
 726
\bibitem{}
 Zinnecker, H. \& Preibisch, T. 1994, A\&A, 292, 152

\end{thebibliography}
\end{document}